\documentclass[useAMS,usenatbib]{mn2e}

\usepackage{epsfig,graphicx,mathptm}
\usepackage{times}
\usepackage{natbib}

\newcommand{\gtrsim}{\ga}
\newcommand{\lesssim}{\la} 
\newcommand{\lognlogs}{$\log N$-$\log S$}

\newcommand{\flunits}{erg s$^{-1}$ cm$^{-2}$}

\newcommand{\ypar}{$y$-parameter}

\newcommand{\lcdm}{$\Lambda$CDM}
\newcommand{\omegam}{$\Omega_{\rm m}$}
\newcommand{\omegal}{$\Omega_{\Lambda}$}
\newcommand{\fnl}{$f_{\rm NL}$}

\newcommand{\wmap}{{\it WMAP}}
\newcommand{\wmapone}{{\it WMAP-1}}
\newcommand{\wmapthree}{{\it WMAP-3}}

\newcommand{\rosat}{{\it ROSAT}}
\newcommand{\erosita}{{\it eROSITA}}
\newcommand{\spt}{{\it SPT}}
\newcommand{\xmm}{{\it XMM-Newton}}

\def\prd{Phys. Rev. D}
\def\aap{A\&A}
\def\apj{ApJ}

\def\apjl{ApJ}
\def\mnras{MNRAS}
\def\araa{ARA\&A}

\def\physrep{Phys. Rep.}

\def\apjs{ApJS}
\title[X-ray and SZ imprints of primordial non-Gaussianities] 
{Imprints of primordial non-Gaussianities in X-ray and SZ signals 
from galaxy clusters}
\author[M. Roncarelli et al.]
{M. Roncarelli$^1$,
L. Moscardini$^{2,3}$, 
E. Branchini$^4$, 
K. Dolag$^{5}$, 
M. Grossi$^{5}$,
F. Iannuzzi$^{5}$ \newauthor
and S. Matarrese$^{6,7}$
\\
$^1$ Centre d'Etude Spatiale des Rayonnements, CNRS/Universit\'e de 
Toulouse, 9 avenue du Colonel Roche, BP44346, 31028 Toulouse 
Cedex 04, France \\
(mauro.roncarelli@cesr.fr) \\
$^2$ Dipartimento di Astronomia, Universit\`a di Bologna, via Ranzani
 1, I-40127 Bologna, Italy (lauro.moscardini@unibo.it)\\
$^3$ INFN -- Sezione di Bologna, 
viale Berti Pichat 6/2, I-40127 Bologna, Italy \\
$^4$ Dipartimento di Fisica, Universit\`a di Roma TRE, 
via della Vasca Navale 84, I-00146, Roma, Italy (branchin@fis.uniroma3.it) \\
$^5$ Max-Planck-Institut f\"ur Astrophysik, Karl-Schwarzschild Strasse
1, D-85741 Garching bei M\"unchen, Germany \\ 
(kdolag,margot,iannuzzi@mpa-garching.mpg.de)\\
$^6$Dipartimento di Fisica ''G. Galilei", Universit\`a di Padova, via
Marzolo 8, I-35131, Padova,
Italy (sabino.matarrese@pd.infn.it)\\
$^7$ INFN -- Sezione di Padova, via Marzolo 8, I-35131, Padova, Italy \\
}

\begin{document}

\date{Accepted ???. Received ???; in original September 2009}

\pagerange{\pageref{firstpage}--\pageref{lastpage}} \pubyear{2009}

\maketitle

\label{firstpage}

\begin{abstract}

    Several inflationary models predict the possibility that the
    primordial perturbations of the density field may contain a degree
    of non-Gaussianity which would influence the subsequent evolution
    of cosmic structures at large scales.  In order to study their
    impact, we use a set of three cosmological DM-only simulations
    starting from initial conditions with different levels of
    non-Gaussianity: \fnl$=0,\pm100$. More specifically, we focus on
    the distribution of galaxy clusters at different redshifts and,
    using suitable scaling relations, we determine their X-ray and SZ
    signals.  Our analysis allows us to estimate the differences in the
    log$N$-log$S$ and log$N$-log$Y$ due to the different initial
    conditions and to predict the cluster counts at different
    redshifts expected for future surveys (\erosita\ and \spt).  We
    also use a second set of simulations assuming a different
    cosmological scenario to estimate how the dependence on \fnl\ is
    degenerate with respect to other parameters.  Our
    results indicate that the effects introduced by a realistic amount
    of primordial non-Gaussianity are small when compared to the ones
    connected with current uncertainties in cosmological parameters,
    particularly with $\sigma_8$. However, if future surveys will be
    associated with optical follow-up campaigns to determine the
    cluster redshift, an analysis of the samples at $z > 1$ can
    provide significant constraints on \fnl. In particular we predict
    that the \spt\ cluster survey will be able to detect $\sim1000$
    clusters at $z>1$ for the Gaussian case, with a difference of
    15--20 per cent associated to \fnl$=\pm100$.

\end{abstract}


%%%%%%%%%%%%%%%%%%%%%%%%%%%%%%%%%%%%%%%%%%%%%%%%%%%%%%%%%%%%%%%%%%%
%%%%%%%%%%%%%%%%%%%% Introduction %%%%%%%%%%%%%%%%%%%%%%%%%%%%%%%%%
%%%%%%%%%%%%%%%%%%%%%%%%%%%%%%%%%%%%%%%%%%%%%%%%%%%%%%%%%%%%%%%%%%%

\section{Introduction} \label{sec:intro}

The universally accepted scenario for the formation of cosmic
structures in the universe is based on the mechanism of gravitational
instability, which assumes that the density fluctuations generated at
some early epoch grow by accreting mass from the surrounding regions
through gravitational processes. The origin of these cosmological
seeds is generally related to the final phases of the inflationary
expansion for which a large variety of theoretical models exist in the
literature \citep[see, e.g.,][for recent
reviews]{kinney08,langlois08,baumann08,baumann09}.  These models
originate perturbations having different statistical properties,
usually investigated in terms of probability distribution function
(PDF) and correlation functions/power polyspectra. In particular, the
most standard slow-rolling models, where a single field is responsible
for the inflationary accelerated expansion, produce fluctuations
having almost uncorrelated phases.  For this reason in cosmological
studies it is usual to assume that the primordial perturbations are
Gaussianly distributed, which leads to the further simplification that
their complete description is possible using the power spectrum
only. However, even the simplest inflationary models allow for small
departures from Gaussianity, which can become more significant in 
non-standard models, like the scenarios based on the curvaton,
the inhomogeneous reheating and the Dirac-Born-Infeld inflation
\citep[see][and references therein]{bartolo04}.  As a consequence, the
observational determination of the amount of non-Gaussianity present
in the primordial fluctuations is now considered not only a general
probe for the inflationary concept, but also a powerful discriminatory
test between its various theoretical models.

Given the infinite variety of possible non-Gaussian models, it is
necessary to introduce a simple way to quantify the level of
primordial non-Gaussianity.  In the recent years it has become 
standard practice to
adopt the dimensionless non-linearity parameter \fnl\ \citep[see][]{salopek90,gangui94,verde00,komatsu01}, measuring the
importance of the quadratic term in a sort of Taylor expansion of the
Bardeen's gauge-invariant potential $\Phi$, namely
\begin{equation}
\Phi=\Phi_{\rm L}+f_{\rm NL} (\Phi^2_{\rm L}-\langle \Phi^2_{\rm L}\rangle)\ ;
\end{equation}
here $\Phi_{\rm L}$ represents a Gaussian random field. In particular, in
this paper we follow the so-called large-scale structure (LSS)
convention, where $\Phi$, that on scales smaller than the Hubble radius
corresponds to the usual Newtonian peculiar potential (but with
changed sign), is linearly extrapolated to the present epoch.

For many years, the study of the statistical properties of the cosmic
microwave background (CMB) has been considered the most efficient way
to measure \fnl. In fact its temperature fluctuations are directly
related to the density perturbations in a regime in which the
non-linearities originated by the subsequent process of gravitational
instability are not modifying their primordial characteristics,
including the PDF.  Different statistical estimators have been applied
to the most recent data, keeping improving constraints on \fnl.
So far the more stringent results come from the analyses of the
5-years {\it WMAP} dataset: assuming a local shape for non-Gaussianity (as
we will do in this paper), \cite{komatsu09} found that \fnl varies
between -12 and 145, while \cite{smith09} 
found $-5<f_{\rm NL}<104$\footnote{We multiplied by 
a factor of 1.3 the original results to convert them 
to the \fnl\ LSS-convention adopted here.}\citep[see also the 
positive detection of non-Gaussian features reported by][]{yadav08}.

More recently it became clear that the LSS represent an alternative tool,
potentially as valid as the CMB to constrain \fnl. In fact, deviations
from an initial Gaussianity induce a different timing for the whole
process of structure formation, providing an interesting framework
to look for specific non-Gaussian imprints. In general, as already
evident from the results of the first generation of non-Gaussian
$N$-body simulations in the early '90s \citep{messina90,
  moscardini91,weinberg92}, if the primordial density field is
positively (negatively) skewed, the formation is favored
(disfavoured) and structures of a given mass form at earlier (later)
epochs. However, there is an important difficulty in using LSS for
constraining \fnl: the late non-linear evolution introduces
additional non-Gaussian features that need to be disentangled from
the primordial ones.  For this reason it has been necessary to have
resort to suitable high-resolution $N$-body simulations to follow the 
growth of the LSS also in the full non-linear regime, and to calibrate
the expected signatures as a function of the primordial
non-Gaussianity. This has been extensively done in the recent years by
different groups \citep[see,
e.g.,][]{kang07,grossi07,dalal08,viel09,desjacques09,pillepich08,grossi09}. The
results allowed to assess the power of LSS as independent probe for
\fnl, in particular they showed that the most evident non-Gaussian
signatures are present in the mass function and clustering (bias and
bispectrum) of dark matter mass haloes \citep[see also the analytical
predictions made
by][]{matarrese00,loverde08,afshordi08,carbone08,mcdonald08,
maggiore09,lam09,valageas09,verde09}.
The first attempts of an application to real data gave very
encouraging constraints: \cite{slosar08}, combining the bias
measurements for two samples of luminous red galaxies and quasars,
found $f_{\rm NL}=37^{+42}_{-57}$, while \cite{afshordi08}, studying
the integrated Sachs-Wolfe effect (ISW) in the NVSS survey, derived $f_{\rm
  NL}=272\pm 127$; error bars are at 2-$\sigma$ level. Notice that in
both cases we report the values as revised by \cite{grossi09} to
include a correction mimicking the ellipsoidal collapse.

Being at the top of the hierarchy of structure formation, galaxy
clusters are in principle ideal probes for constraining \fnl. 
Indeed, the statistics of fare events, either galaxy clusters or 
deep voids \citep{viel09} are very sensitive to primordial non-Gaussianity. 
However, until now the use of clusters as non-Gaussian probes has been 
hampered by two practical problems: first, the
observational estimate of their mass is affected by large
uncertainties, whatever are the method and the observational band
adopted; second, it is difficult to build up samples that span a large 
range of redshift and are also statistically complete. The first
problem can be in some way overcome by using the 
scaling relations
(expected from theoretical arguments and confirmed by observations)
existing between mass and different observables (see below). The
second problem will be solved in the next years, thanks to the efforts
of a set of (in progress or planned) surveys, which promise to provide
large samples of galaxy clusters covering a volume comparable to the
horizon size: see, e.g., DES, PanSTARRS, BOSS, LSST, ADEPT, EUCLID.

In this paper we will focus on estimating the signals produced by 
galaxy clusters
in the X-ray band and through the Sunyaev-Zel'dovich 
\citep[SZ,][]{sunyaev72,sunyaev80} effect in 
different non-Gaussian scenarios. We will take advantage of
the fact that, in both cases, the corresponding observables, i.e.  the
X-ray luminosity and the Compton \ypar, are related to the mass by
well calibrated scaling relations. This will allow us to introduce
selection criteria mimicking the characteristics of specific
surveys. In particular we will consider the properties of the wide
surveys planned with \erosita\ \citep{predehl07} and
\spt\ \citep{carlstrom09}, as examples of future X-ray and SZ projects,
respectively. The main goal of this work is to figure out what are the
observational evidences of the presence of some level of primordial
non-Gaussianity, discussing the possibility of constraining \fnl\ with
these future datasets.  Notice that the same samples have been
considered for a similar work in \cite{sefusatti06,oguri09,fedeli09}. 
In this paper we address this problem using numerical rather than
analytical tools. The advantage is twofold. First of all, $N$-body
simulations permit to fully account for non-linear evolution which,
instead, is usually accounted for by analytical models in
an approximate way only. Second, numerical experiments allow us 
to extract realistic mock cluster catalogs that can be easily used 
for modeling the observational selection, which is more 
difficult to account for in a Fisher matrix-like approach.

The plan of this work is as follows. In Section~\ref{sec:models} 
we present the
numerical simulations of non-Gaussian models on which the following
analysis is based; we also describe the method applied to construct
the light cones. Section~\ref{sec:xray} introduces our model for the 
X-ray emission from galaxy clusters and reports the corresponding 
results in terms of number counts, paying attention to the expected 
results for the \erosita\ wide survey.  Section~\ref{sec:tsz} is 
devoted to the model for the SZ signal and to the corresponding 
results, given in terms of source counts and statistical properties 
of the maps; the specific case of the \spt\ wide survey is treated. 
Section~\ref{sec:diff} discusses the
possibility of using the differential redshift distribution to
constrain the primordial \fnl. Finally, in Section~\ref{sec:conclu} 
we draw our conclusions.

%%%%%%%%%%%%%%%%%%%%%%%%%%%%%%%%%%%%%%%%%%%%%%%%%%%%%%%%%%%%%%%%%%%
%%%%%%%%%%%%%%%%%%%% Models and method %%%%%%%%%%%%%%%%%%%%%%%%%%%%
%%%%%%%%%%%%%%%%%%%%%%%%%%%%%%%%%%%%%%%%%%%%%%%%%%%%%%%%%%%%%%%%%%%

\section{Models and method} \label{sec:models}

\subsection{The simulation sets} \label{ssec:simul}

In order to study the possibility of detecting the signatures of
primordial non-Gaussianity in the LSS of the Universe we must take
into account the complete process of structure formation. For this we
make use of the outputs of two different sets of DM-only simulations,
focusing on the distribution of the DM haloes associated to the galaxy
clusters, as identified in the different snapshots.

The first simulation set \citep[presented in][]{grossi09} consists of
three DM-only simulations performed with different levels of
primordial non-Gaussianity, that, expressed in terms of the
dimensionless non-linearity parameter \fnl, are \fnl=0, $\pm$100; the
case \fnl=0 corresponds to the standard Gaussian case. The initial
conditions were set by assuming a flat \lcdm\ model dominated by a
cosmological constant with parameters chosen to be consistent with the
\wmap\ three-year results \citep[][\wmapthree\ hereafter]{spergel07}:
namely, the density contributions of cosmological constant and matter
correspond to \omegal =0.76 and \omegam =0.24, respectively, while the
normalization of the power spectrum of density fluctuations is fixed
as $\sigma_8$=0.8, being $\sigma_8$ the r.m.s. matter fluctuation into
a sphere of radius $8 h^{-1}$ Mpc.  The three simulations started from
the same random generation of initial conditions with the only
difference consisting in the different value of \fnl. All the runs,
carried out with the $N$-body code \textsc{gadget-2}\
\citep{springel01,springel05}, followed the evolution of $960^3$ DM
particles inside a cubic volume of $1200 h^{-1}$ Mpc per side, with
each particle having a mass of $m = 1.4 \times 10^{11} h^{-1}
M_\odot$; here $h$ represents the Hubble parameter defined as $h\equiv
H_0/$(100 km s$^{-1}$ Mpc$^{-1}$)=0.7. The gravitational force has
been computed using a Plummer equivalent  
softening length $\epsilon = 25 h^{-1}$ kpc.

These simulations produced 14 outputs in the redshift range 
$0 \le z \le 4$. For
each snapshot we produced a catalogue of DM haloes identified using a
{\it friends-of-friends} (FOF) algorithm, adopting a linking length of
0.2 times the mean interparticle distance: with this choice the size
of the haloes roughly corresponds to their virial mass, $M_{\rm vir}$.
Since in this work we are interested in the X-ray and SZ signals
produced by galaxy groups and clusters, we only considered DM haloes
having $M_{\rm vir} > 10^{13} h^{-1} M_\odot$.  For each halo we kept
the information on the mass and the position of its center inside
the cosmological volume.

The second set of simulations is the one described in
\cite{grossi07}. It consists on 7 runs covering a wider range of
non-Gaussianity: \fnl=0, $\pm$100, $\pm$500, $\pm$1000. The main
differences with respect to the first set are that the box size is
only $500 h^{-1}$ Mpc with $800^3$ particles and that the cosmological
model adopted is close to a \wmap\ first-year \citep[][\wmapone\
hereafter]{spergel03} cosmology, with the following cosmological
parameters: \omegal =0.7, \omegam =0.3, $\sigma_8$=0.9. For this set
we have 21 outputs in the range $0 \le z \le 4$ and a halo catalogue
for each snapshot obtained in the same way as described before and
with the same mass limit.  Notice that in this paper this last set of
simulations will be mainly used to test the impact of different
cosmological parameters compared to non-Gaussianity: for this reason 
we will discuss only the results of the most extreme and the Gaussian 
model (\fnl$=0,\pm 1000$). This simulation set will also be used in 
order to check the effect of the finite box size on our results.

\subsection{Constructing the light-cones} \label{ssec:lcones}

As said in Section \ref{sec:intro}, in
order to study the impact of primordial non-Gaussianity on the LSS we
adopt an observationally-oriented approach. To this purpose we use the
halo catalogues described in Section \ref{ssec:simul} to produce 
mock light-cones by stacking several simulation volumes. In
particular we want to cover the redshift range $0 \leq z \leq 4$,
which corresponds to a comoving distance of 5249 $h^{-1}$ Mpc (5019
$h^{-1}$ Mpc with the cosmology adopted for the second set). This
length requires to stack 5 (11 for the second set)
times the simulation box. However, in order to obtain a better
redshift sampling, we divide the simulation volume into slices along
the line of sight. The number of slices varies from cube to cube and
their comoving distance intervals are created in order to allow us to 
use all of the 14 (21 for the second set) snapshots. More precisely,
for any given distance from the observer we compute 
the corresponding time elapsed from the big bang and we choose 
the snapshot that better approximates this value.

In order to avoid the repetition of the same structures along the line
of sight, for every stacked simulation volume we perform a
randomization of the halo spatial coordinates: for every cube we
randomly choose the axis to put along the line of sight, we assign a
50 per cent probability to reflect each axis and, since our
simulations assume periodic boundary conditions, we proceed to a
random recentering of the coordinates. In order to preserve the whole
information on the structures inside the simulations' volume, the
slices belonging to the same cube undergo the same randomization
process.  With this method, which is similar to the one adopted by
\cite{roncarelli06a}, we obtain 18 different slices
belonging to 5 independently randomized cubes (31 and 11 for the
second simulation set, respectively). This process is repeated with
the same initial random seed for all the simulations of the set.  Each
light-cone produced in this way spans an angle of 13.1 deg (5.71 deg
for the second set) per side, determined by the length of the box at
the maximum redshift, $z=4$.

By varying the initial random seeds we can obtain different light-cone
realizations that we can use to assess the statistical robustness of
our results.  For each non-Gaussian model, we created 20 (100)
different light-cone realizations, thus covering a total area of 3432
deg$^2$ (3260 deg$^2$). However, it is important to note that this
area cannot be considered as completely independent, being produced
starting from the same finite volume of the simulation: as a
reference, at $z \simeq 4$ the same simulation volume is completely
stacked in all realisations, while at $z \simeq 0.5$ we are able to
produce about 15 independent volumes crossing the light-cones.

\begin{figure*}
\includegraphics[width=0.90\textwidth]{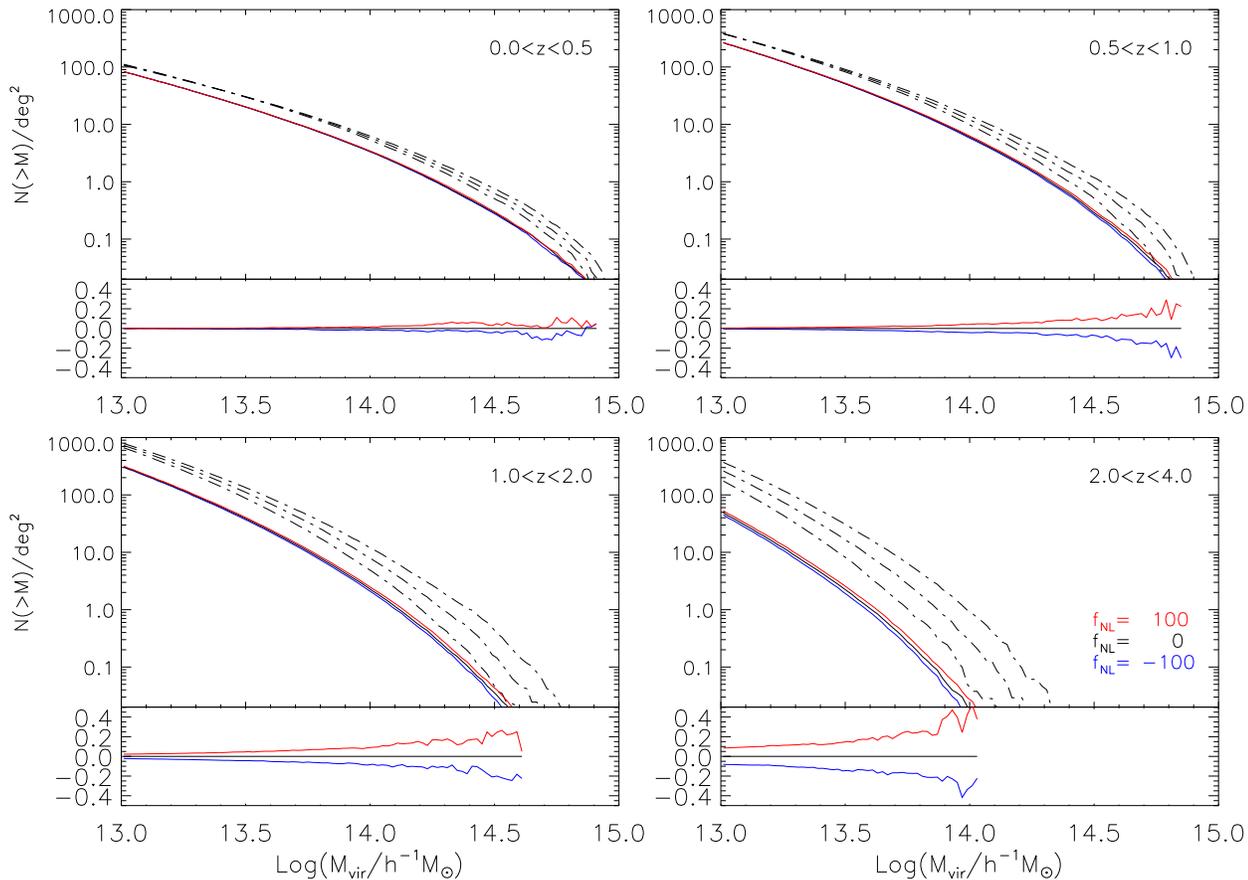}
\caption{ The $\log N$-$\log M$ for the two simulation sets in
  different redshift bins. In the upper panels the solid lines refer
  to the first simulation set (average of 20 light-cone realizations)
  for the three different levels of non-Gaussianity: \fnl=0,$\pm$100. 
  The dot-dashed lines refer to the second simulation
  set (average of 100 light cones, \wmapone\ cosmology) for models
  with \fnl=0, $\pm$1000.  In the lower panels we show, for the first
  set of simulations only, the difference $\Delta N/N$, computed
  with respect to the Gaussian (\fnl=0) simulation. }
\label{fig:mass_funct}
\end{figure*}

Using  the whole set of light-cone realizations we compute the mass 
functions, expressed in terms of number of objects per solid 
angle, for 4 redshift bins. We show in Fig. \ref{fig:mass_funct} 
the results for the three models of the first set 
(\fnl=$0,\pm100$) and for the \fnl=$0,\pm1000$ models of the 
second set.

Looking at the $0 \le z \le 0.5$ bin,  
the three mass functions of the first set are basically 
indistinguishable. The 
non-Gaussian deviations become more evident in the tail at 
higher masses and at higher redshifts: in the second bin ($0.5 
\le z \le 1$) the \fnl$=\pm100$ models introduce a $\pm 5$ per 
cent difference in the high mass ($M_{\rm vir} \gtrsim 
10^{14} h^{-1} M_\odot$) 
cluster counts, which increases to about 10 per cent at $z > 1$.
This is in agreement with the fact the modification of the
distribution of the primordial density fluctuations primarily affects
the formation of the biggest haloes at early epochs, as already 
discussed in \cite{grossi09}.

When comparing the two simulation sets, it is clear 
that the differences
between the various non-Gaussian scenarios are very small 
compared to the ones resulting from the change of the cosmological 
model. 
For example, even at $z>1$ the \fnl$=100$ model of the first set 
adds only about 10 per cent counts to the Gaussian case 
($M_{\rm vir}>10^{14} h^{-1} M_\odot$), while the 
\wmapone\ cosmology scenario predicts more than three times as 
much objects. This fact highlights that the uncertainties with 
cosmological parameters, particularly with $\sigma_8$, are critical 
when addressing the problem of detecting primordial non-Gaussianities 
in the LSS. More in the detail, analytical models of the mass 
functions \citep[e.g.][]{sheth02} predict that a difference of 
$\pm0.01$ in $\sigma_8$ produces a change of about 10 per cent 
in the total integrated counts ($M_{\rm vir}>10^{14} h^{-1} M_\odot$), 
raising to about 15 per cent at $z>1$, where the \fnl\ parameter is 
expected to produce the most significant effects. This aspect will be 
considered in the following analyses.

It is worth to note also that for masses higher than 
$\approx 10^{14.5} h^{-1} M_\odot$ the mass
functions of the second simulation set steepen and approach the
first-simulation ones: this is an artificial effect produced by the
smaller box size of the second simulation (500 instead of
$1200 h^{-1}$ Mpc). In fact this causes a loss of power for
perturbations at large scales, corresponding to a lower abundance of
objects with high masses, and gives an indication of the range of
validity of our simulation sets.

%%%%%%%%%%%%%%%%%%%%%%%%%%%%%%%%%%%%%%%%%%%%%%%%%%%%%%%%%%%%%%%%%%%
%%%%%%%%%%%%%%%%%%%%%%% Modeling X-ray counts %%%%%%%%%%%%%%%%%%%%%
%%%%%%%%%%%%%%%%%%%%%%%%%%%%%%%%%%%%%%%%%%%%%%%%%%%%%%%%%%%%%%%%%%%

\section{Modeling X-ray counts} \label{sec:xray}

Since our simulations consider only DM particles, in order to compare
our results with present and future cluster surveys we need to define
a model to associate the baryonic component to each DM halo.  In
particular we will focus here on modeling the X-ray emission (and the
SZ signal, described in Section~\ref{sec:tsz}), exploiting observed
and predicted scaling relations.  As said in Section \ref{ssec:simul},
our method provides the virial mass $M_{\rm vir}$ of our clusters,
while often scaling relations are published using other mass
definitions (like $M_{200}$, $M_{500}$ and so on): therefore, in order
to convert $M_{\rm vir}$ to the required value, we assume that our DM
haloes follow a NFW density profile \citep{navarro97} with a
concentration parameter $c$ given by the $c(M,z)$ relation proposed by
\cite{dolag04}.  Notice that here we explicitly neglect the presence
of any diffuse emission from the IGM associated to the warm-hot
intergalactic medium (WHIM) and concentrate only on signals coming
from galaxy clusters and groups. Even if the WHIM component is
expected to contribute significantly to the total LSS signal, its
presence is not expected to significantly affect the clusters count
rate \citep[see, e.g.,][] {roncarelli06a}.

In order to associate an X-ray luminosity to a DM halo we assume 
the phenomenological mass-luminosity relation found by 
\cite{stanek06}, namely
\begin{equation}
\frac{L_X}
{h_{70}^{-2} \, 10^{44} \, {\rm erg \, s^{-1}}} = 
L_{15,0} \, E(z)^s 
\left( \frac{M_{200}}{10^{15} h^{-1} M_\odot} \right) ^p \, ,
\label{eq:l-m}
\end{equation}
where $L_X$ is the luminosity in the [0.1-2.4 keV] band and $M_{200}$
is the mass of the cluster inside a radius enclosing 200 times the
critical density $\rho_c(z)$ of the Universe at the redshift of the
cluster. The normalization $L_{15,0}$ corresponds to the luminosity of
an object with $M_{200}=10^{15} h^{-1} M_\odot$ at $z=0$. The term
$E(z)$ represents the redshift evolution of the Hubble parameter,
\begin{equation}
E(z)=\Omega_{\rm m}(1+z)^3+\Omega_\Lambda \,
\label{eq:m-l}
\end{equation}
while the value of $s$ is assumed to correspond to the self-similar
evolution case, $s=7/3$. The best-fit parameters $L_{15,0}$ and $p$ 
depend on the assumed cosmological model. \cite{stanek06} publish their
results assuming both \omegam=0.24 ($\ln L_{15,0}=1.19$,
$p=1.46$) and \omegam=0.30 ($\ln L_{15,0}=1.34$, $p=1.59$),
therefore we can take those values
as a reference for our first and second simulation sets, respectively.
We also take into account a scatter in the $L_{\rm X}-M$ relation that
we fix to 17 per cent, as measured by \cite{reiprich02}.

In this work we will determine the observed clusters fluxes in the
[0.5-2 keV] and [0.5-5 keV] bands to compare our results with the
abundances derived from \rosat\ X-ray clusters survey \citep{rosati02}
and to predict the expected counts for the forthcoming \erosita\
survey \citep{predehl07}, respectively.  In order to calculate the
band corrections, we need to assume the ICM temperature that
determines the spectral distribution of the emitted radiation. For
this purpose we consider our haloes as isothermal and we use the
$M_{200}-T_{200}$ relation which \cite{arnaud05} obtained from a
sample of ten nearby clusters observed with \xmm, adding a
self-similar redshift evolution. Notice that their sample covers the
temperature range 2--9 keV, so we are forced to extend this relation
to smaller objects, where a steepening of the $M_{200}-T_{200}$ 
relation is expected. In fact, \cite{arnaud05} obtain a slope higher 
of $\sim0.2$ when restricting their sample only the hottest 
($T>3.5$ keV) clusters. Even if neglecting this effect may lead to
an underestimate of the cluster temperatures for the small objects,
we checked that it does not have a strong impact on our final results. 
For example, if we assume a further steepening of $0.2$ in colder 
($T<2$ keV) clusters, this leads to a change of about 1 per cent in 
the faint-end of the \lognlogs, and the relative differences 
between the non-Gaussian scenarios remain unchanged.

It is known that the ICM cools mainly via {\it bremsstrahlung}
emission which is the main physical process responsible of the X-ray
emission of galaxy clusters. Therefore we can model the emission of
the gas assuming a free-free spectrum with a Gaunt factor $g(E/k_{\rm
  B})T_{200}=(E/k_{\rm B})^{-0.3}$ \citep[see, e.g.,][ for more
details]{borgani99}. With this simplifying assumption we are
neglecting the presence of other known emission processes like
line-emission from metals, which can give a non-negligible
contribution especially to low-temperature ($T \lesssim 2$ keV)
clusters; however, for the reasons explained above,
neglecting this process has no significant impact on our final
results.

Once obtained the band correction $f_{\rm band}$, we calculate the
cluster flux in a given band as
\begin{equation}
S_{\rm band} = \frac{L_X f_{\rm band}}{4 \pi d_L(z)^2} \, ,
\end{equation}
where $d_L(z)$ is the luminosity distance of the cluster.

\begin{figure}
\includegraphics[width=0.50\textwidth]{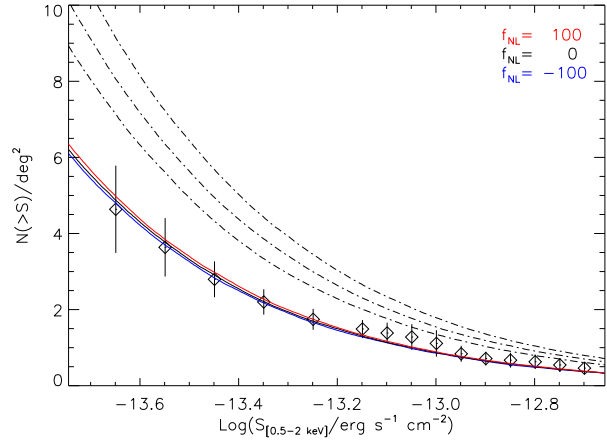}
\caption{The number of haloes as a function of the flux limit in the
  [0.5-2 keV] band for the two simulation sets.  The solid lines refer
  to the three models of the first set (average of 20 light-cone
  realizations) with \fnl=0,$\pm$100. The dot-dashed line refer to the
  second simulation set (average of 100 light-cones, \wmapone\
  cosmology) for models with \fnl=0, $\pm$1000.  The diamonds with
  errorbars correspond to the number counts derived by 
  \protect \cite{rosati02} from \rosat\ X-ray clusters
  survey. }
\label{fig:check_counts}
\end{figure}

Fig.~\ref{fig:check_counts} shows the number counts as a function of
the X-ray flux limit in the [0.5-2 keV] band for the three models of
the first set and for the \fnl=0,$\pm$1000 of the second simulation
set. Our results are compared with the counts obtained from the
\rosat\ survey, in the same band. All of the first three models show a
good agreement with the data, having error bars much wider than the
difference associated to the change of the \fnl\ parameter.  On the
contrary, the counts obtained with the \wmapone\ cosmology are
significantly higher (even when considering the standard Gaussian
case): the only way to fit the data with these cosmological parameters
would be to assume a strong negative evolution of the luminosity with
redshift, e.g.  adding an extra factor $(1+z)^\beta$ to equation
\ref{eq:l-m}, with $\beta \simeq -2$, which is absolutely unrealistic 
\citep[see, e.g.,][]{ettori04}. 
These results show that the dependence
of the cluster counts on different cosmological parameters, and
particularly on $\sigma_8$, is much higher with respect to the one on
the value of \fnl: this highlights the difficulty of constraining the
level of the primordial non-Gaussianity with the current uncertainties
in the cosmological parameters.

\begin{figure*}
\includegraphics[width=0.90\textwidth]{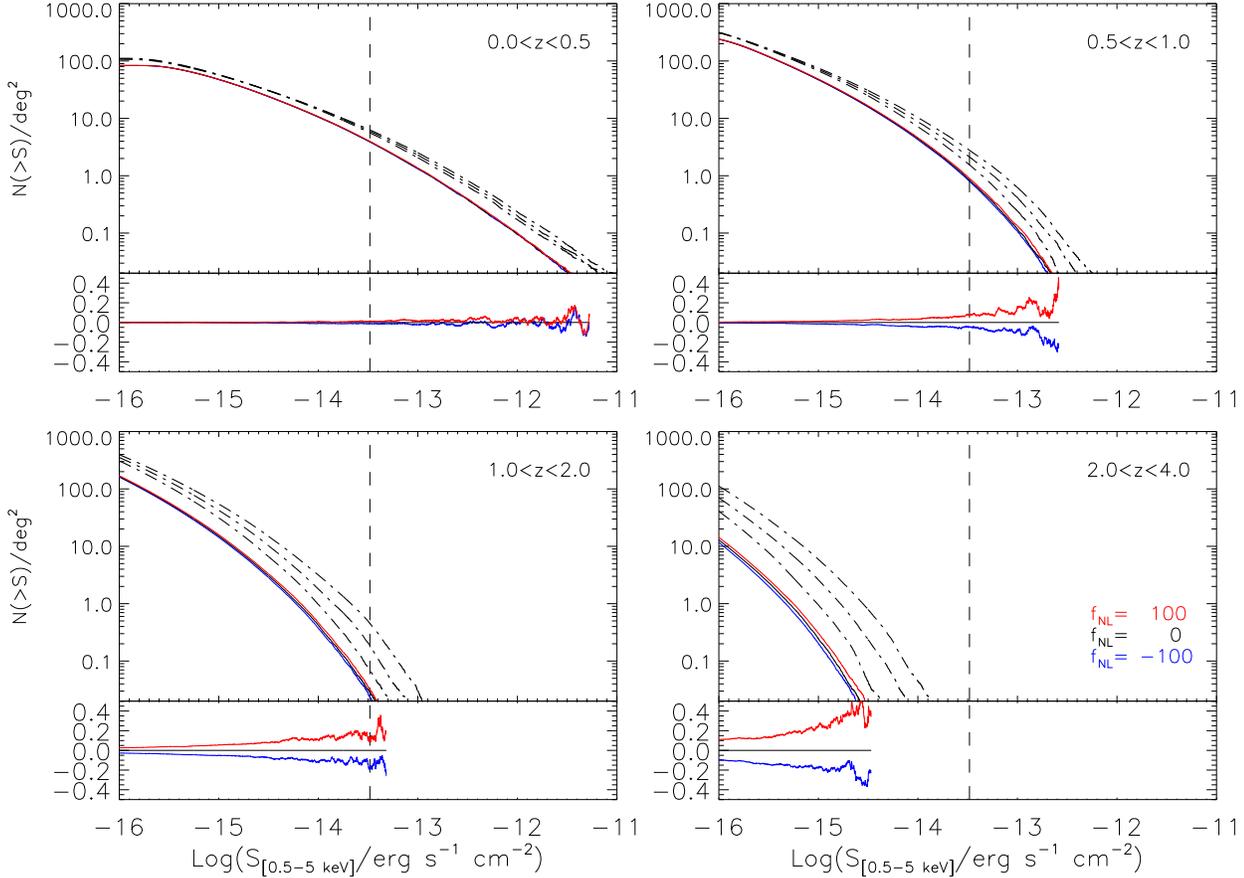}
\caption{As in Fig. \ref{fig:mass_funct}, but for the $\log N$-$\log S$
  in the [0.5-5 keV] band. The dashed vertical line indicates the
  \erosita\ flux-limit, corresponding to $3.3 \times 10^{14}$
  \flunits. }
\label{fig:logn-logs}
\end{figure*}

Fig. \ref{fig:logn-logs} shows the \lognlogs\ in the [0.5-5 keV] band
in different redshift intervals. Again, the number counts differences
between the different non-Gaussian scenarios are about one order of
magnitude smaller than the corresponding differences obtained by
increasing the value of $\sigma_8$ from 0.8 to 0.9. Anyway, it is
interesting to note that in the redshift interval $0.5<z<1$, the
number counts expected assuming the \erosita\ detection limit ($3.3 
\times 10^{-14}$ \flunits) differ by about 10 per cent when assuming
\fnl$=\pm100$ with respect to the Gaussian model.  These relative
differences grow to about 20 per cent in the $1<z<2$ interval and
increase for higher fluxes: this result is in agreement with the
expected evolutionary scenario emerging from
Fig.~\ref{fig:mass_funct}, with higher mass (and then more luminous)
haloes forming earlier in models with positive \fnl.

Table \ref{tab:erosita} shows the number of galaxy clusters with
$M_{200} > 10^{14} h^{-1} M_\odot$ that are expected to be detected by
the \erosita\ wide survey under our assumptions. For our predictions
we assume an effective area of 20000
deg$^2$ and a detection limit of 3.3$\times 10^{-14}$ \flunits\ in the
[0.5-5 keV] band. For the Gaussian model we predict $\sim 60000$
detections, one per cent of them at redshift larger than unity.  The
changes due to the presence of a moderate level of primordial
non-Gaussianity (\fnl$=\pm100$) are small, but always larger than the
expected poissonian error: considering the total counts, the variation
is about 4 per cent, while for the redshift bin $1<z<2$ the abundances
change by 15 per cent.  However, once again, the dependence on \fnl\
is much smaller when compared to the one on the power spectrum
normalization $\sigma_8$: at $z>1$, 
the differences in the cluster counts associated to the 
\fnl$=\pm100$ models are comparable to an uncertainty of $\pm 0.01$ 
in $\sigma_8$ (see the discussion in Sect.~\ref{ssec:lcones}). For what 
concerns the total integrated counts, the number of objects 
detectable by the \erosita\ survey predicted by the Gaussian 
simulation of the second set ($\sigma_8=0.9$) grows by a
factor of 2: $\sim$120000 objects, with a similar redshift
distribution.  This confirms the necessity of having alternative
derivations of the main cosmological parameters to allow to exploit
the power of cluster counts to constrain the level of primordial
non-Gaussianity \citep[see also][]{fedeli09}.

\begin{table}
\begin{center}
  \caption{ The number of detected haloes (with $M_{200} > 10^{14} h^{-1}
    M_\odot$) predicted for the \erosita\ wide survey in different
    redshift bins.  The quoted errors are assumed to be poissonian.}
\begin{tabular}{rccccc}

               &  & \multicolumn{4}{c}{Number counts}  \\
$f_{\rm NL}$ &  & $0<z<0.5$ & $0.5<z<1$ & $1<z<2$ & Total  \\
\hline

 -100 & & 41590$\pm$204 & 15934$\pm$126 & 478$\pm$ 22 & 58007$\pm$241 \\
    0 & & 42318$\pm$206 & 16715$\pm$129 & 571$\pm$ 24 & 59610$\pm$244 \\
  100 & & 43443$\pm$208 & 18015$\pm$134 & 641$\pm$ 25 & 62105$\pm$249 \\

\hline
\label{tab:erosita}
\end{tabular}
\end{center}
\end{table}

%%%%%%%%%%%%%%%%%%%%%%%%%%%%%%%%%%%%%%%%%%%%%%%%%%%%%%%%%%%%%%%%%%%
%%%%%%%%%%%%%%%%%%%%%%%%% The thermal SZ effect %%%%%%%%%%%%%%%%%%%
%%%%%%%%%%%%%%%%%%%%%%%%%%%%%%%%%%%%%%%%%%%%%%%%%%%%%%%%%%%%%%%%%%%

\section{The thermal SZ effect} \label{sec:tsz}

Another important observable quantity to study galaxy clusters is 
the thermal SZ (tSZ) effect, namely the inverse-Compton scattering 
of the CMB photons caused by the electrons present in the hot 
intracluster plasma \citep[see, e.g.,][for detailed reviews]{birkinshaw99,
carlstrom02}. This effect causes a distortion in the CMB blackbody 
spectrum, whose intensity in a given direction can be expressed in 
terms of the Compton \ypar\, defined as 
\begin{equation}
y \equiv \frac{k_{\rm B}\, \sigma_T}{m_e\, c^2} \int{n_e T_e \, dl} \ ;
\label{eq:ypar}
\end{equation}
here $k_{\rm B}$ is the Boltzmann constant, $\sigma_T$ is the 
Thompson cross section, $m_e$ is the electron mass, $c$ is the light speed, 
while $n_e$ and $T_e$ represent the electron number density and temperature,
respectively. The
distortion of the signal results in a difference $\Delta T$ in the
measured temperature which depends on the observational frequency. In
the Rayleigh-Jeans (RJ) limit this is given by
\begin{equation}
\frac{\Delta T}{T_{\rm CMB}} = -2 y \ ,
\label{eq:dt_t}
\end{equation}
where $T_{\rm CMB}$=2.726 K is the CMB temperature 
\citep{mather94}, 
thus producing a temperature decrement, which can be as high 
as $\Delta T \approx 10^{-3}$ K for the central regions of the 
most massive clusters.

Following an observationally oriented approach, the interesting quantity 
to be evaluated for a given halo is 
the integrated Compton \ypar\, $Y$, defined as 
\begin{equation}
Y \equiv \int_\Omega y \, d\Omega =
\frac{1}{d_A^2(z)} 
\left( \frac{k_{\rm B} \sigma_{\rm T}}{m_e c^2} \right)
\int_V n_e T_e \, dV \, ,
\end{equation}
where $\Omega$ is the solid angle subtended by the cluster 
and $V$ is its physical volume. This adimensional quantity depends on 
the angular diameter distance $d_A(z)$, and constitutes an equivalent 
to the flux in the X-rays.  Therefore it is useful to introduce the 
intrinsic Compton \ypar\, defined as
\begin{equation}
Y^{\rm int} \equiv Y d_A(z)^2 \, ,
\end{equation}
which is roughly proportional to the mass and to the temperature of
the object. Using hydrodynamical simulations it has 
been possible to calibrate scaling relations between the 
cluster mass and its SZ observables 
\citep[see, e.g.,][]{diaferio05,shaw08}: here we adopt the 
$M-Y^{\rm int}$ relation, found by \cite{nagai06}, that can be
expressed in the form
\begin{equation}
\frac{Y^{\rm int}_{200}}{{\rm Mpc}^2} = 
A_{14} \times 10^{-6} E(z)^s 
\left( \frac{M_{200}}{10^{14} h^{-1} M_\odot} \right)^\alpha \, ,
\end{equation}
where the pedex indicates that we are considering quantities computed
inside the volume enclosed by $R_{200}$. By fitting the data computed 
from their simulated clusters sample, \cite{nagai06} obtained
$A_{14}=2.56$ and $\alpha=1.70$ (we consider the results of their 
CSF simulation, which accounts for a variety of 
physical processes 
of the baryonic component): we used
this relation assuming, as we did for the X-ray modeling, a
self-similar redshift evolution, corresponding for the tSZ effect to 
$s=2/3$.

\begin{figure*}
\includegraphics[width=0.90\textwidth]{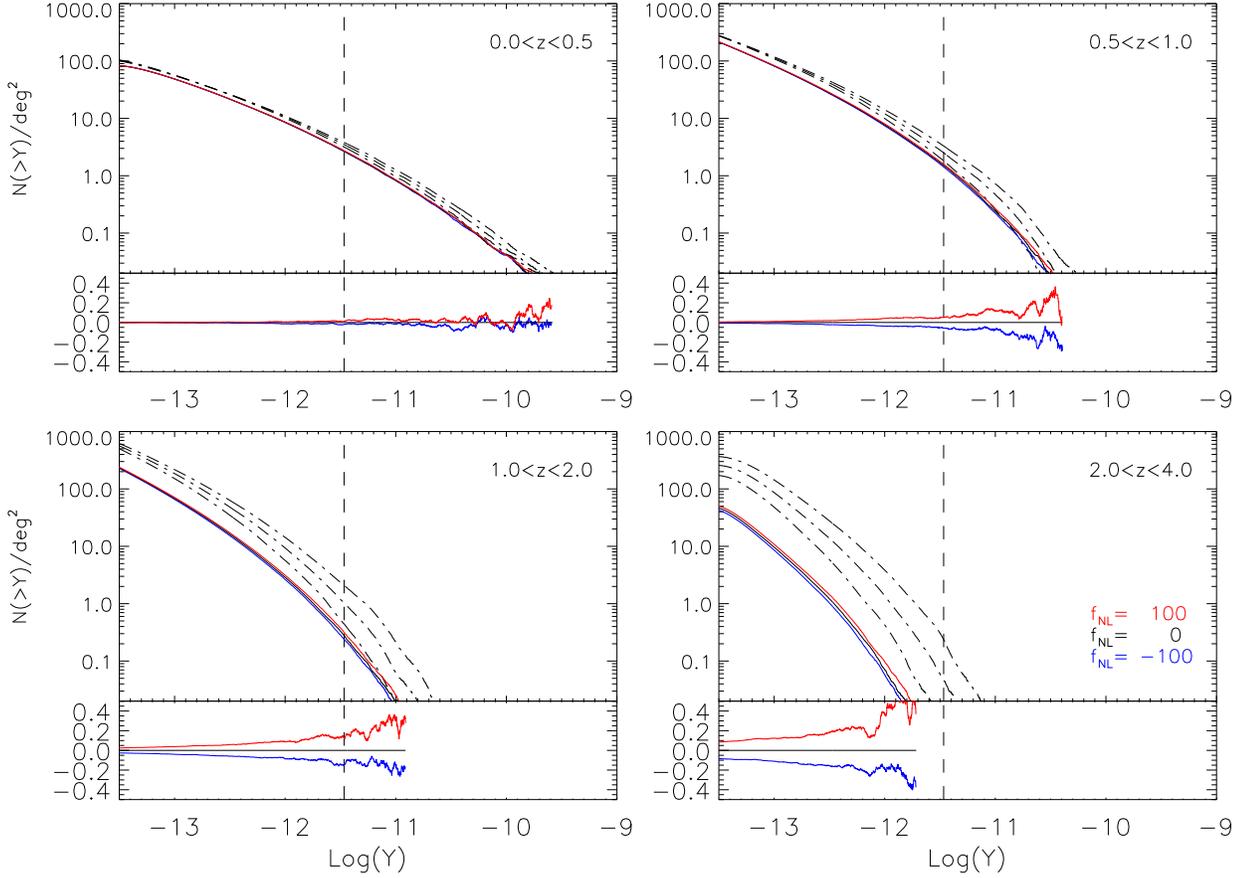}
\caption{ As in Fig. \ref{fig:mass_funct}, but for the
  $\log N$-$\log Y$.  The dashed vertical line indicates the \spt\
  flux-limit, i.e. 5 mJy at 150 GHz, corresponding to $Y \sim 3.4 \times
  10^{-12}$. 
 }
\label{fig:logn-logy}
\end{figure*}

We show the results of our $\log N$-$\log Y^{\rm int}$ in Fig.
\ref{fig:logn-logy}. Most of the conclusions derived from the analysis
of the X-ray results apply also for the tSZ effect: the uncertainties
in the estimate of $\sigma_8$ make it very challenging to discriminate
between different non-Gaussian models. However it is worth to notice
that, since the tSZ effect is not affected by redshift dimming as the
X-ray flux, at higher redshifts ($z>1$) the number of detections is
still significant. For example, assuming the flux limit expected for
the \spt\ survey \citep[i.e. 5 mJy at 150 GHz, corresponding to 
$Y \simeq 3.4 \times 10^{-12}$, see][]{majumdar03},
we predict the 
possibility of detecting $\sim 0.3$ high-$z$ objects per square
degree: given the expected area of 4000 deg$^2$, this leads to more than one
thousand objects, enough to potentially discriminate the $\sim20$ per
cent difference in the cluster counts predicted for the \fnl$=\pm100$
models. Note that these deviations are comparable to an 
uncertainty of about $\pm0.02$ in the primordial power spectrum 
normalization $\sigma_8$.
The expected number counts from the \spt\ survey in different
redshift ranges are reported in Table~\ref{tab:spt}.

\begin{table}
\begin{center}
  \caption{ The number of detected haloes (with $M_{200} > 10^{14}
    h^{-1} M_\odot$) predicted for the \spt\ wide survey in
    different redshift bins.  The quoted errors are assumed to be
    poissonian.  }
\begin{tabular}{rccccc}

               &  & \multicolumn{4}{c}{Number counts}  \\
$f_{\rm NL}$ &  & $0<z<0.5$ & $0.5<z<1$ & $1<z<2$ & Total  \\
\hline

 -100 & & 7796$\pm$88 & 5961$\pm$ 75 &  960$\pm$ 31 & 14456$\pm$120 \\
    0 & & 7958$\pm$89 & 6041$\pm$ 78 & 1124$\pm$ 34 & 15137$\pm$123 \\
  100 & & 8144$\pm$90 & 6357$\pm$ 80 & 1268$\pm$ 36 & 15784$\pm$126 \\

\hline
\label{tab:spt}
\end{tabular}
\end{center}
\end{table}

\subsection{Integrated properties of the tSZ signal}

\begin{figure*}
\includegraphics[width=0.42\textwidth]{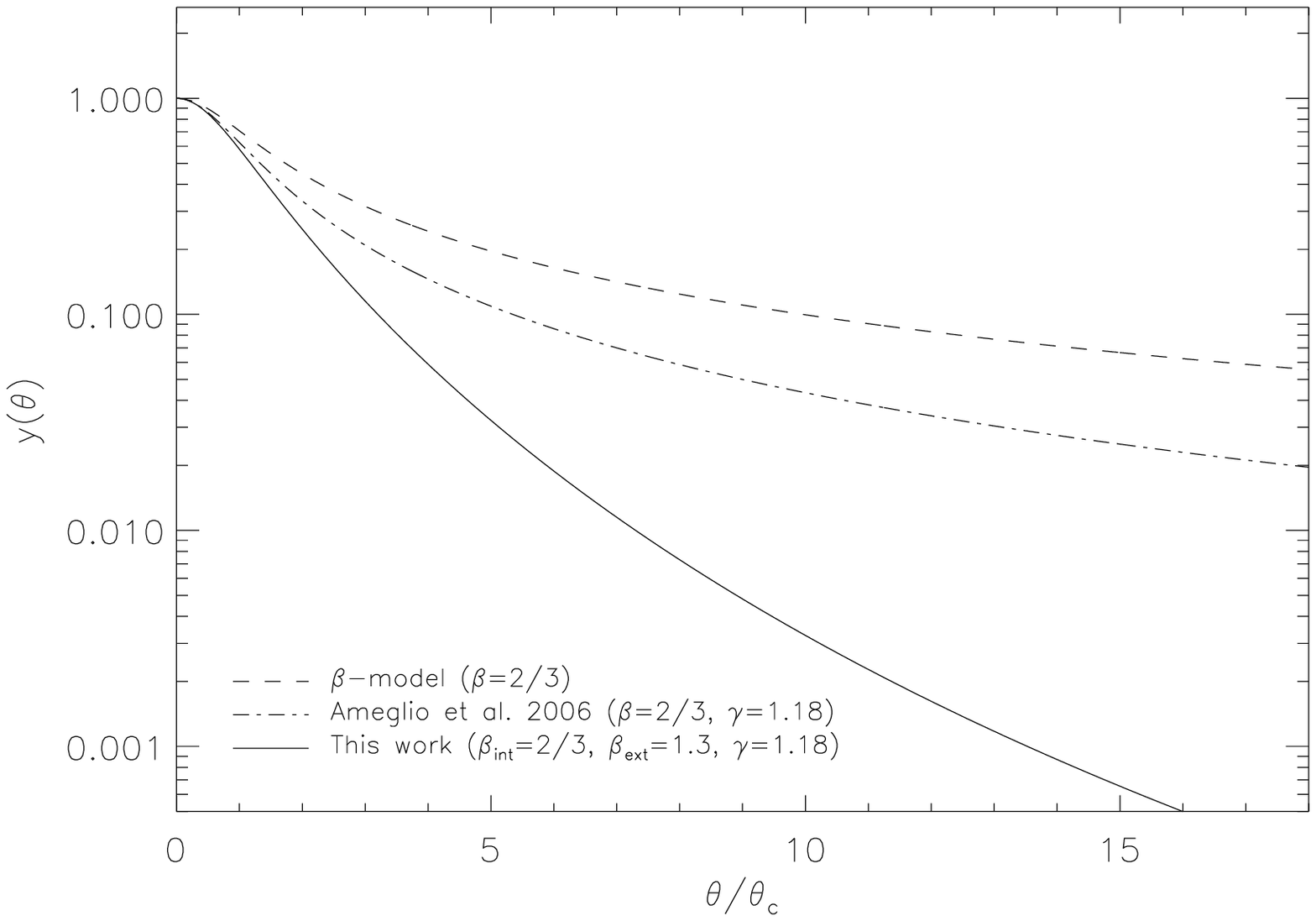}
\includegraphics[width=0.42\textwidth]{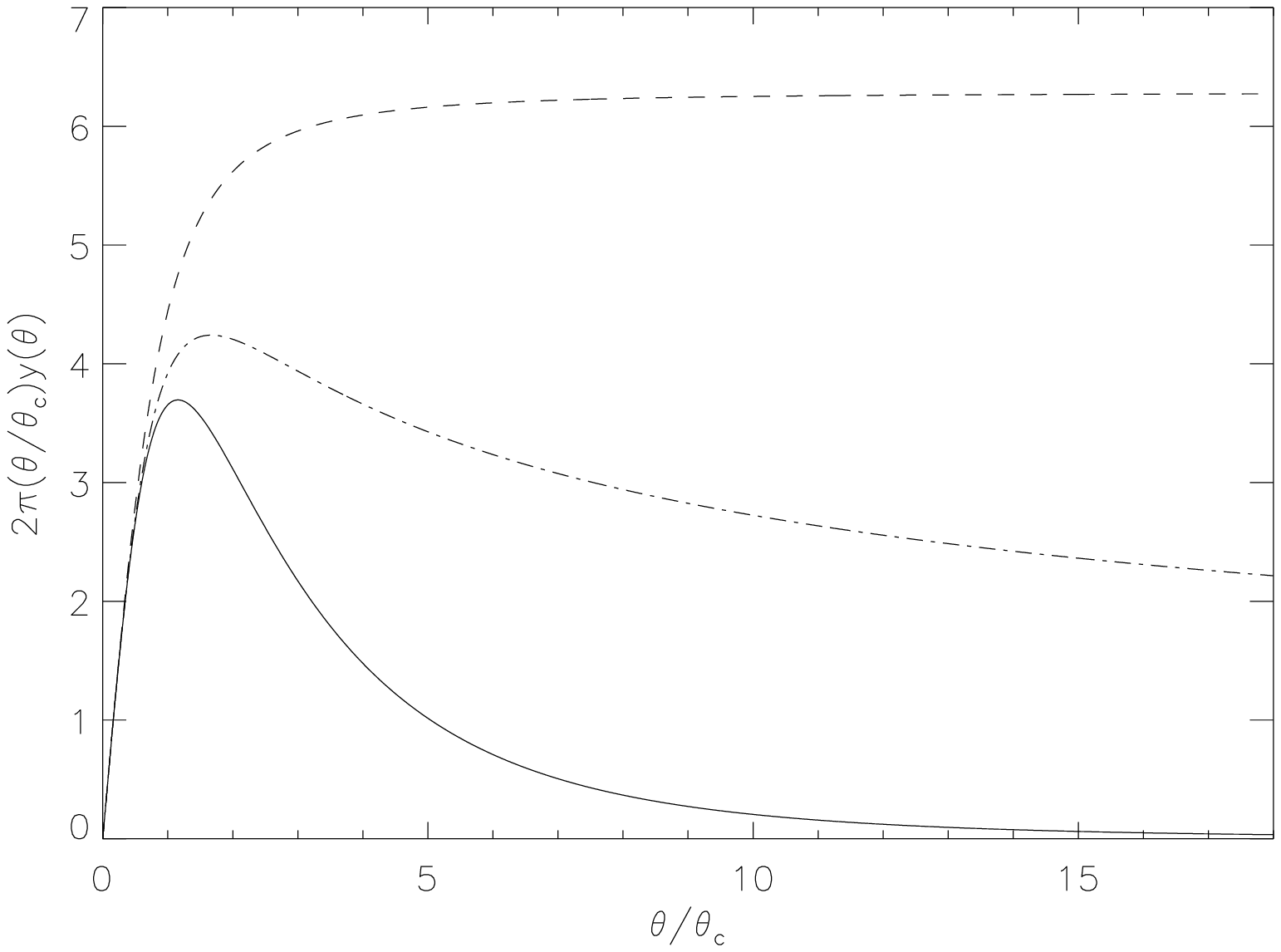}
\caption{ Left panel: comparison between the Compton \ypar\ profiles
  for an isothermal $\beta$--model (dashed line), the polytropic
  $\beta$--model proposed by \protect \cite{ameglio06} (dot-dashed
  line) and the rolling-$\beta$ polytropic profile (solid line) used
  in this work. Right panel: distribution of the signal as a function
  of the distance from the centre for the same three profiles.}
\label{fig:prof_comp}
\end{figure*}

Apart from cluster counts, other global 
quantities like the average Compton \ypar\ and the tSZ power spectrum 
can be affected by the presence of a non-Gaussian signature in 
the primordial power spectrum of perturbations.
In order to study these observables, it is necessary to
create and analyse mock maps of the \ypar.
Since galaxy clusters are extended
sources, we need to make further assumptions about the density and
temperature profiles of the haloes to model the distribution of the
signal on the sky. Since the tSZ signal receives a significant
contribution also from the external regions of galaxy clusters, the
modelisation must take into account the steepening of the slope of
these profiles in the regions around the virial radius \citep[see,
e.g.,][]{vikhlinin06,roncarelli06b}. In this context, it is important
to note that a classic $\beta$-model \citep{cavaliere78} would fail
simply because it does not converge for large distances from the
centre, even when assuming a decrease of the temperature with a
polytropic index, as adopted in \cite{ameglio06}.  Therefore, we start
from the suggestion of \cite{ameglio06} and slightly modify their
results by introducing a rolling-$\beta$ polytropic profile for the
tSZ signal, namely
\begin{equation}
y(\theta) = y_0 \left[ 1+\left( \frac{\theta}{\theta_{\rm c}} \right)^2 \right] ^ 
            {-3\beta_{\rm eff}(\gamma-1)/2} \ ,
\end{equation}
where $\theta$ is the angular separation from the cluster centre, 
$\theta_{\rm c}$ is the angular size of the core radius (assumed to 
be $r_{\rm c}=0.1R_{200}$), $\gamma=1.18$ is the polytropic index of the 
gas. The effective slope $\beta_{\rm eff}$ is given by 
\begin{equation}
\beta_{\rm eff} = -\beta_{\rm ext} \left(
\frac{x+\beta_{\rm int}/\beta_{\rm ext}}{x+1}\right) \ ,
\end{equation}
where $x\equiv\theta/\theta_{\rm c}$ and $\beta_{\rm ext}$ and
$\beta_{\rm int}$ are the external and internal slopes for the density
profile, respectively.  In this way we can tune these two parameters
in such a way that this expression converges to the profile of
\cite{ameglio06} (and to a $\beta$-model) in the inner part, while in
the outer part it steepens to allow the value of the integrated
$Y^{\rm int}$ to converge to a finite value ($\gamma \, \beta_{\rm
  ext} > 1$). We choose $\beta_{\rm int}=2/3$ and $\beta_{\rm
  ext}=1.3$, where the latter value is taken in agreement with the
analyses of \cite{roncarelli06b} on the density profiles in the
outskirts of simulated galaxy clusters. With 
this choice of parameters, the external part of our $y(\theta)$ 
profile also agrees with the results of \cite{haugboelle07}.
A visual comparison of
these different profiles for Compton \ypar\ is shown in 
the left panel of Fig.~\ref{fig:prof_comp}: while the three profiles 
converge to the same value at the centre, the rolling-$\beta$ 
polytropic profile is significantly lower already at $\sim3 \theta_{\rm c}$ 
($\sim 0.3 R_{200}$). When considering the distribution of the signal (right 
panel), it is easy to deduce that only with the profile adopted in this 
work $Y^{\rm int}$ converges to a finite value for high values of 
$\theta$.  It is 
worth to notice that even adopting this profile, the total signal up to 
$R_{200}$ is only the 93 per cent of the total signal: this indicates 
that using any other shallower profile would lead to non-negligible 
biases.

\begin{figure*}
\includegraphics[width=0.28\textwidth]{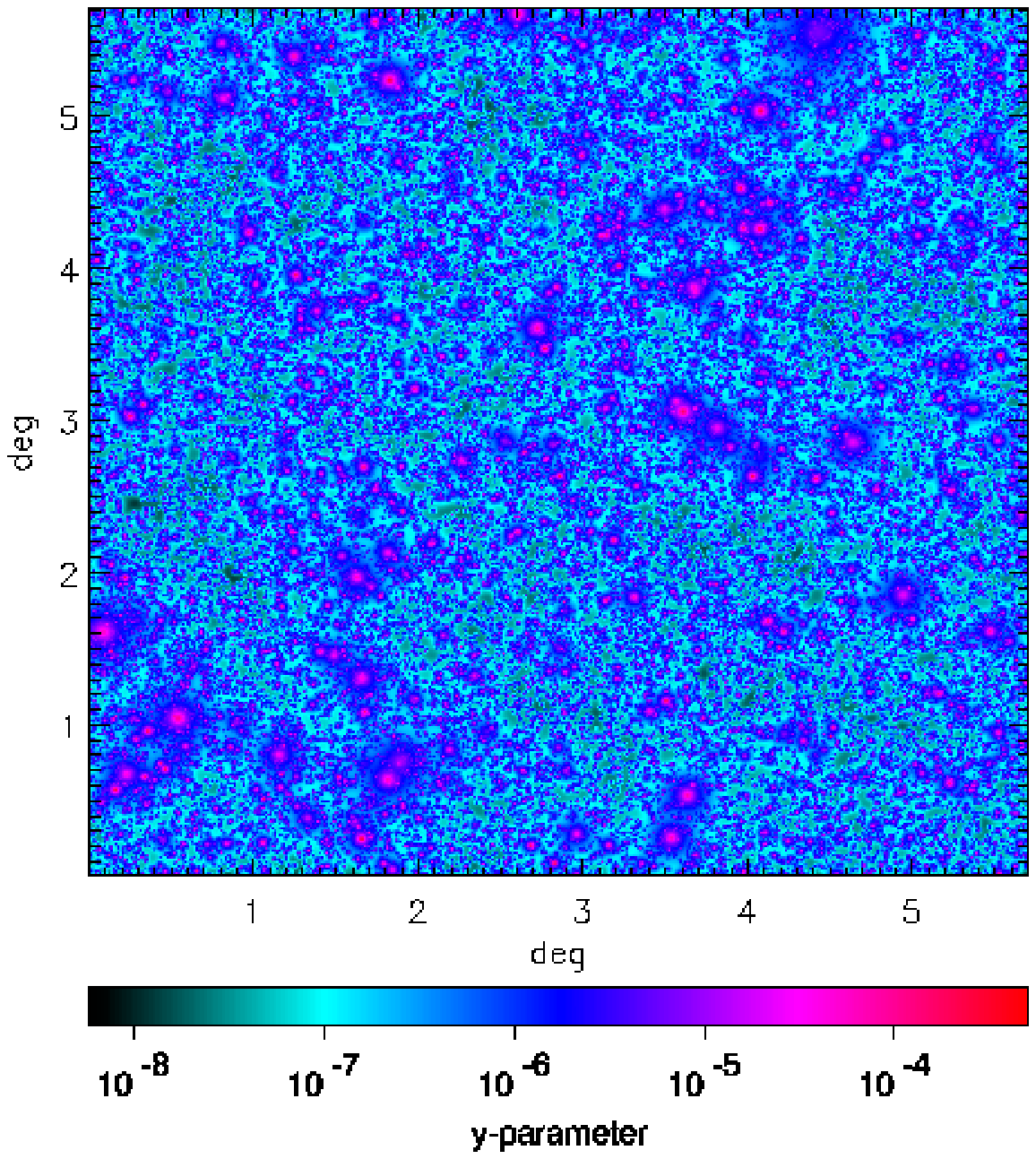}
\includegraphics[width=0.28\textwidth]{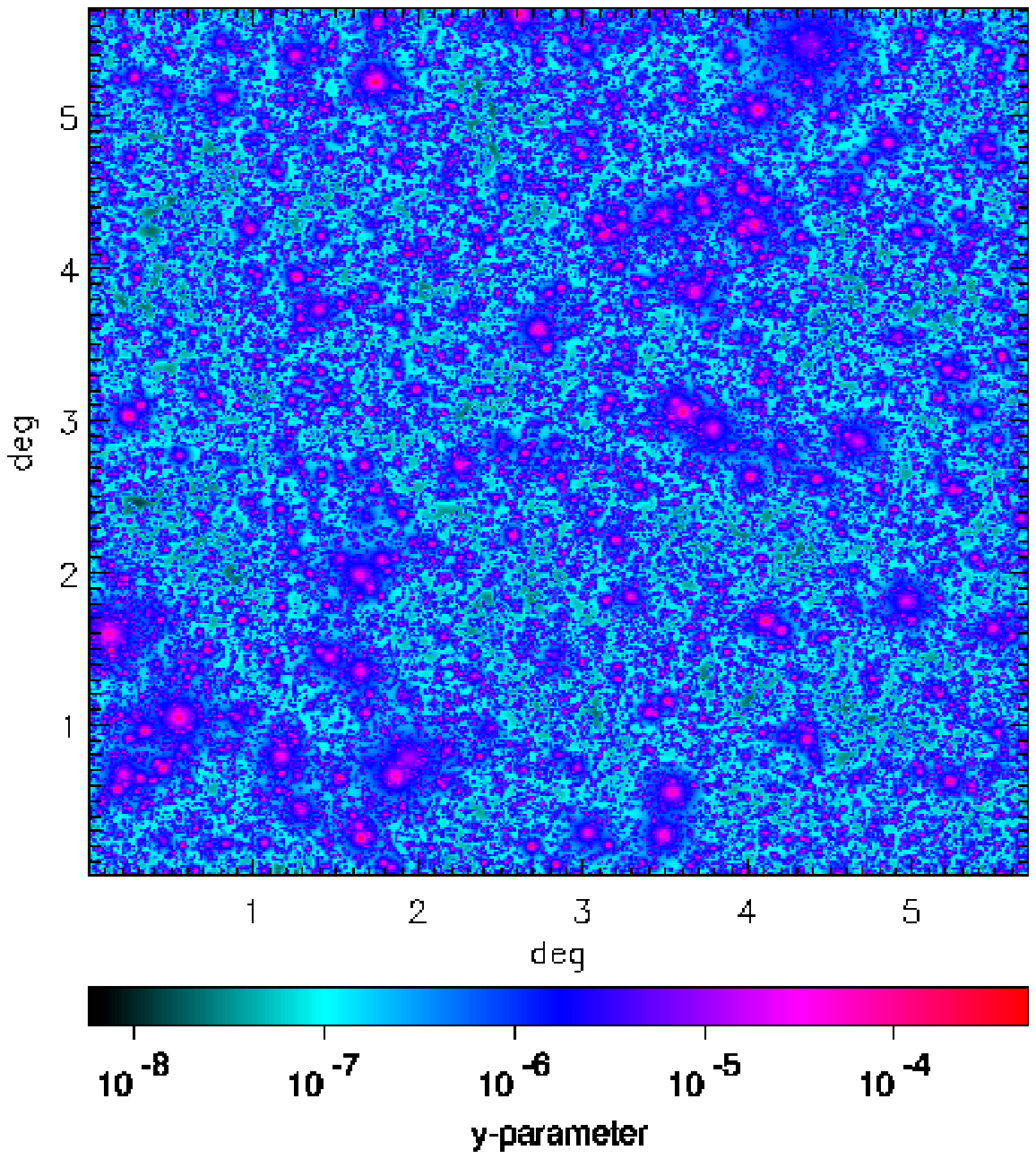}
\includegraphics[width=0.28\textwidth]{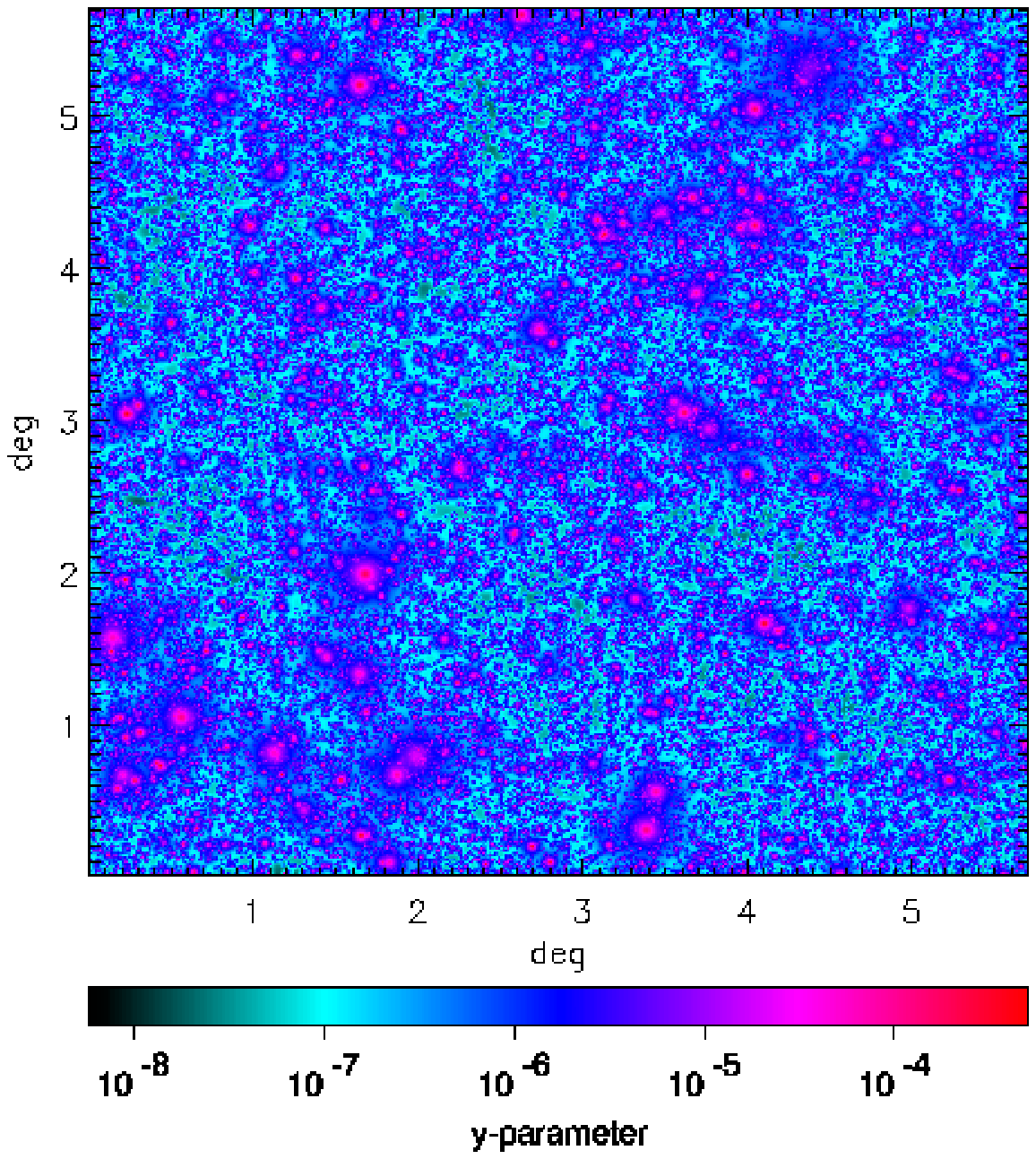} 
\caption{Examples of maps for the tSZ signal (expressed in terms of
  the Compton \ypar) for three different levels of primordial
  non-Gaussianity, $f_{\rm NL}=-1000,0,1000$ (left, center and right
  panels, respectively). These maps have been obtained from the second
  simulation set (\wmapone\ cosmology): they are 5.71$^\circ$ on a
  side with a pixel size of (20 arcsec)$^2$. Notice that the three
  maps refer to the same light-cone realization. }
\label{fig:maps}
\end{figure*}

Then we used this profile as a weight to distribute the total tSZ
signal of each halo into the pixels of our maps. Since the adopted
profile has non-zero contribution even at large scales, we set up the
integration limit to the 99 per cent of the total value: this means
integrating up to $\sim 2.3 r_{200}$.  We follow the procedure
described in \cite{roncarelli06a} to smooth the signal of the SPH
particles and obtain the tSZ map corresponding to each simulated
light-cone realization. At the end our analysis will be based on a
total of 60 maps from the first simulation set and 700 from the second
one.  As an example, in Fig. \ref{fig:maps} we show the Compton \ypar\
maps corresponding to the same light-cone realization, as obtained
from the \fnl$=0,\pm1000$ simulations of the second set. With these
extremely large amounts of primordial non-Gaussianity, it is possible
to recognize the expected behavior: an evidence of more (less)
clustered signal for positive (negative) \fnl, compared to the
Gaussian case.  Notice that the average value of the \ypar\ 
computed over all the maps of the Gaussian simulation with the
\wmapone\ cosmology is $<y>=9.71 \times 10^{-7}$, while considering a
\wmapthree\ cosmology (first set of simulations) this value drops to
$<y>=5.22 \times 10^{-7}$, in good agreement with the expected
scaling, $y \propto \Omega_{\rm m} \sigma_8^{3.5}$ \citep[see, e.g.,][]
{komatsu02,diego04}.
The last figure is also consistent with the results of
\cite{roncarelli07} who, analysing a high-resolution hydrodynamical
simulation \citep{borgani04} based on the same cosmological model,
obtained a value of $<y>=1.19 \times 10^{-6}$ with about half of the
signal originated from the WHIM (not considered in this work).
Notice that the mean values derived for the models with the
\wmapthree\ cosmology but with \fnl$=100$ and \fnl$=-100$ are
$<y>=5.37 \times 10^{-7}$ and $<y>=5.08 \times 10^{-7}$, respectively, 
thus comparable within few percent to the Gaussian case.

\subsection{Angular power spectrum}
\label{ssec:pow_sp}

\begin{figure}
\includegraphics[width=0.50\textwidth]{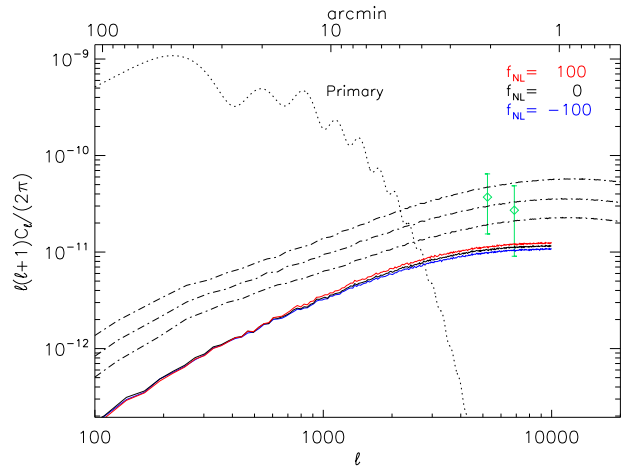}
\caption{ The power spectrum of the tSZ signal in the RJ limit as a
  function of the multipole $\ell$ for the two simulation sets.  The
  solid lines refer to the first simulation set (average of the 20
  light-cone realizations) for the three different levels of
  primordial non-Gaussianity (\fnl=0,$\pm$100), while the dot-dashed
  lines refer to the second simulation set (average of 100
  light-cones, with \wmapone\ cosmology) for \fnl=0,$\pm$1000. The
  dotted line represents the primary CMB signal calculated using
  \textsc{cmbfast} \protect \citep[]{seljak96} adopting the
  \wmapthree\ cosmology. The diamonds with errorbars ($1 \sigma$) 
  represent the 
  measurements of \protect \cite{dawson02} with the {\it BIMA} 
  experiment.}
\label{fig:powersp}
\end{figure}

To characterize the statistical properties of the tSZ effect, it is
important to study its power spectrum at different multipoles, in
particular focusing on the angular scales at which the tSZ signal is
expected to dominate the primary CMB anisotropies ($\ell \gtrsim
2000$). For the complete set of maps generated as described in the
previous Section, we computed the tSZ power spectra, using a method
based on Fast Fourier Transforms, adopting the approximation of flat
sky (given the reduced extension of the maps) and assuming the RJ
frequency limit.  Finally, the corresponding averages are shown in
Fig.~\ref{fig:powersp} for the different models.

Again, when considering the first set of simulations, the differences
between models with various values of \fnl\ are very low when compared
to the variations related to a change of the cosmological scenario:
about 10 per cent in both senses for the \fnl$=\pm 100$ models,
compared to a factor of about 3 when changing $\sigma_8$ and \omegam.
This is expected as, given the dependence $C_\ell \propto \sigma_8^7$, 
the difference in the $\sigma_8$ choice accounts alone
for a factor 2.3, with the remaining difference associated to the
\omegam\ parameter. For these reasons, the perspective of
constraining \fnl\ seems to be quite demanding without an independent
derivation of the main cosmological parameters.

The presence of a possible non-Gaussianity in post-inflationary 
perturbations has been also claimed by \cite{sadeh07} as a possible 
explanation of the anomalous values of the tSZ power spectrum 
obtained by the {\it BIMA} experiment. In particular, \cite{dawson02} 
measured an high value of $\Delta T = 16.6^{+5.3}_{-5.9}$ $\mu$K 
at $\ell = 5237$ \citep[see, however, the smaller estimate obtained by]
[with the {\it SZA} experiment]{sharp09}. According to our results, 
explaining the {\it BIMA} results with primordial non-Gaussian 
fluctuations alone (and keeping $\sigma_8 = 0.8$) 
would lead to values of \fnl\ unrealistically high (\fnl$ \gg 100$). 
On the contrary, if slightly higher values of $\sigma_8$ are considered, 
the tSZ power spectrum would agree within $1 \sigma$ with the result 
obtained by \cite{dawson02}. In this context, a positive value of \fnl\, 
within current upper limits, could also contribute to boost the tSZ 
signal in order to explain these measurements.

\section{Differential redshift counts} \label{sec:diff}

From the results presented above it is clear that a significant
detection of the signatures of a possible primordial non-Gaussianity
based on global properties of galaxy clusters (e.g. number counts,
mass functions, etc.) appears very difficult and well beyond the
possibility of current and planned surveys. The main reason is not only
the degeneracy of the results with other uncertain cosmological
parameters (mostly $\sigma_8$), but also the fact that these
observables are dominated by low-redshift haloes ($z \lesssim 0.5$)
where the LSS properties of the different non-Gaussian models show
smaller differences.

\begin{figure*}
\includegraphics[width=0.42\textwidth]{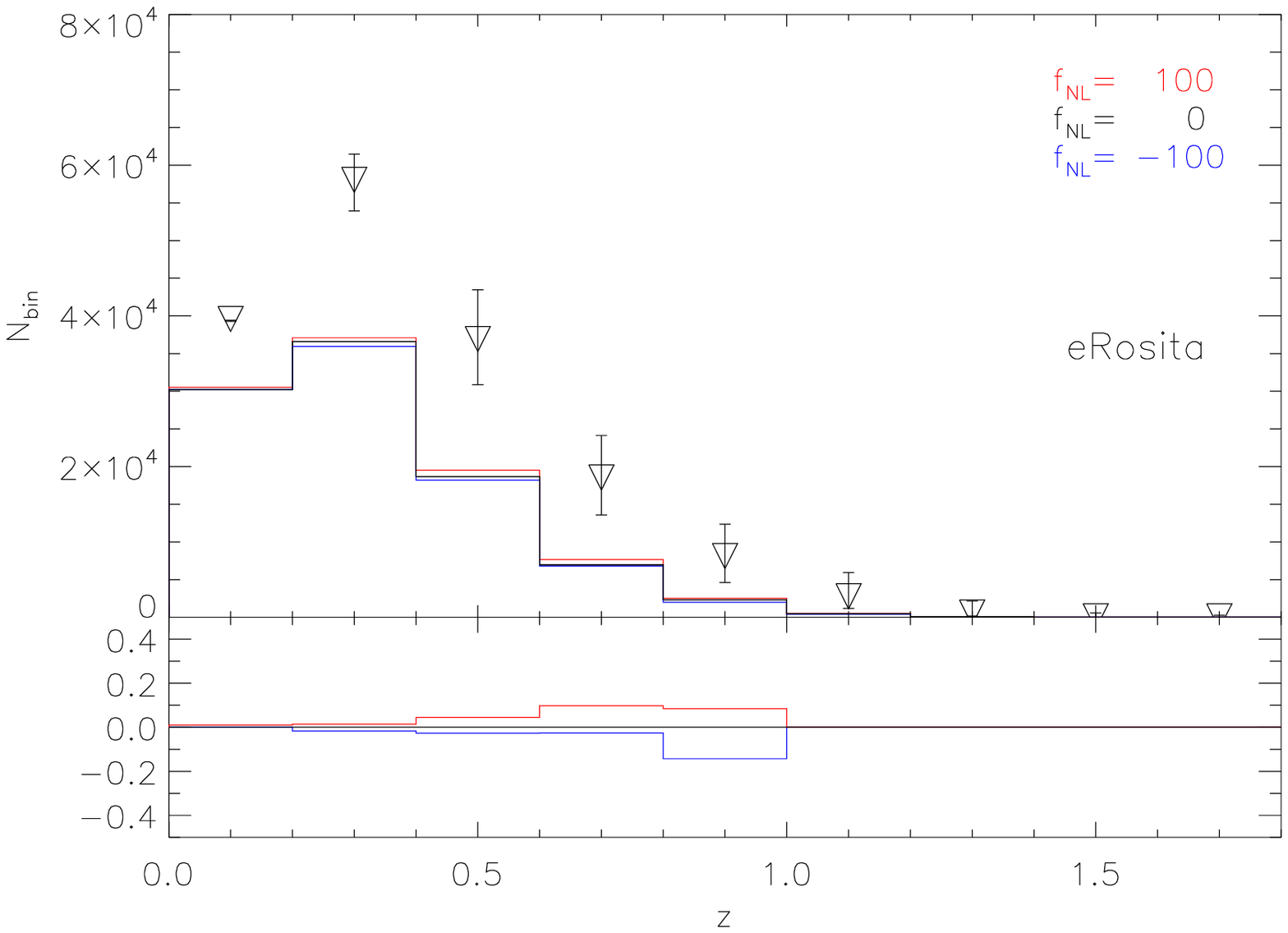}
\includegraphics[width=0.42\textwidth]{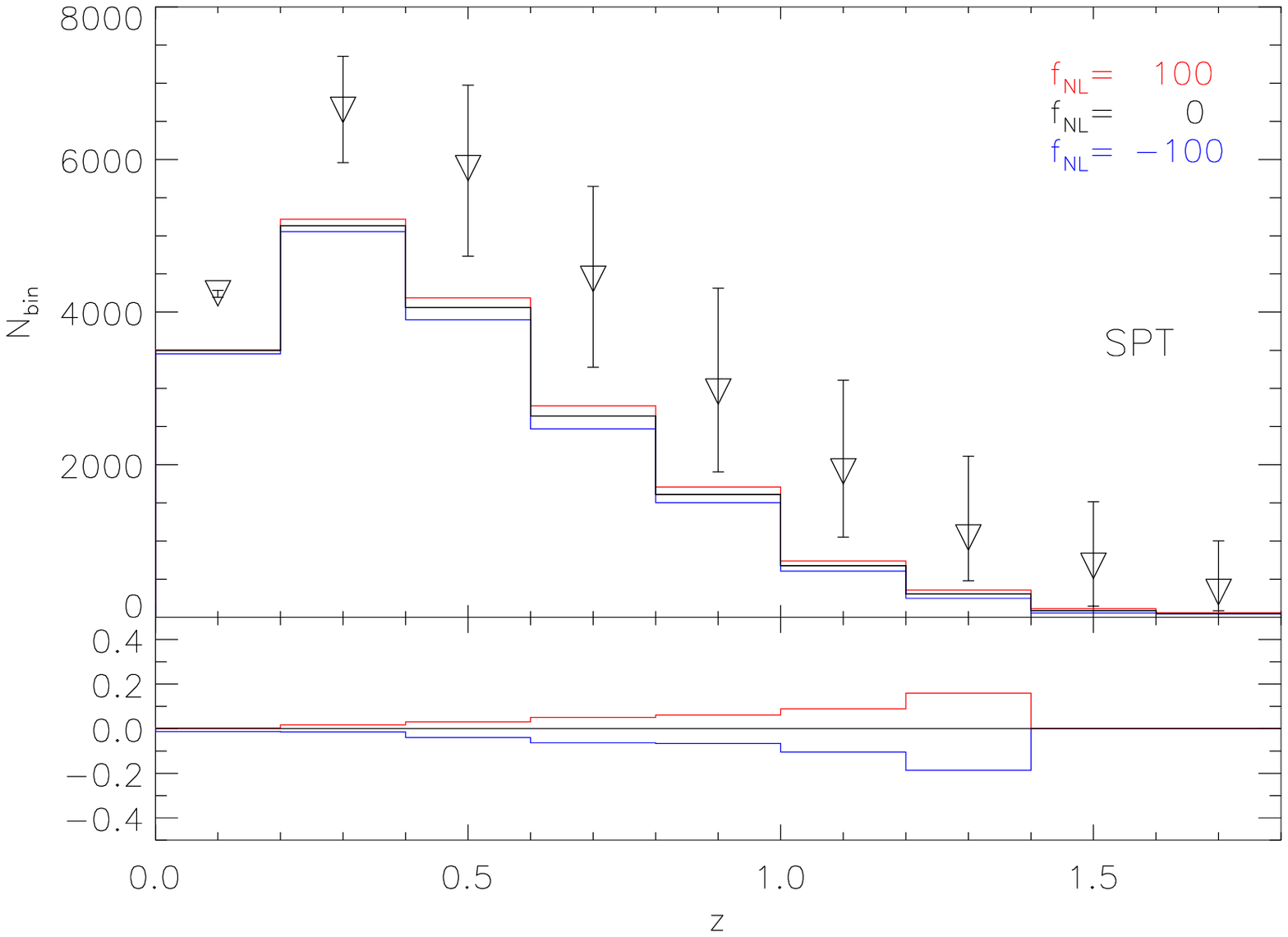}
\caption{The predicted number counts (in different redshift bins with
  $\Delta z =0.2$) of objects detected by \erosita\ (flux limit of
  $3.3 \times 10^{-14}$ \flunits\ in the [0.5-5 keV] band; left panel)
  and \spt\ (flux limit corresponding to $S > 5$ mJy at 150 GHz; right
  panel). The solid lines refer to the first simulation set for the
  three different levels of non-Gaussianity: \fnl=0,$\pm$100.  The
  triangles correspond to the results of the Gaussian simulation of
  the second set, with errorbars indicating how the values change for
  the corresponding non-Gaussian models with \fnl = $\pm 1000$.  In
  the lower panel we show, for the first set only,
  the difference $\Delta N/N$, computed with respect to the Gaussian
  (\fnl=0) simulation.  }
\label{fig:dn_dz}
\end{figure*}

For these reasons the most reasonable strategy to break the degeneracy
between \fnl\ and the other parameters can be a study of the evolution
with redshift of the cluster counts. In Fig.~\ref{fig:dn_dz} we show
for the different models under analysis the redshift distribution of
the objects that will be detected by the \erosita\ and \spt\ cluster
surveys (left and right panels) in the X-ray and tSZ, respectively.

Looking in more detail at the \erosita\ results, the cluster
abundances at $z \gtrsim 0.5$, where the satellite is expected to
detect about 15,000 objects (see Tab.~\ref{tab:erosita}), show a
relative difference of about 10 per cent.  Therefore, in principle,
the possibility of estimating, thanks to a dedicated follow-up
campaign, a high number of redshifts for the objects detected by
\erosita\ at $z>0.5$, would allow to obtain their redshift
distribution, increasing the chance of constraining \fnl, once the
value of $\sigma_8$ is derived using the abundance of low-redshift
objects.

A similar conclusion can be drawn by looking at the redshift
distribution of the \spt\ clusters (right panel). The possibility of
detecting very high-redshift ($z>1$) clusters with an SZ survey is
certainly promising, since in this redshift range the relative
differences grows to $\sim20$ per cent.  However, the possibility of
obtaining a significant amount of redshift estimates for these objects
(which are $\sim 1000$, see Section~\ref{sec:tsz}) is of course much
lower, due to their lower signal.

\section{Conclusions} \label{sec:conclu}

In this work we used a set of cosmological $N$-body simulations to
investigate the impact of primordial non-Gaussianity (parametrised in
terms of \fnl) on the LSS.  From their outputs we constructed halo
catalogues at different redshifts and, making use of suitable scaling
relations between masses and observables, we computed their expected
X-ray emission and SZ signal. In particular we investigated the
possibility of constraining \fnl\ with future projects, namely
\erosita\ and \spt\ cluster surveys.  Moreover we discussed the
degeneracy with other uncertain cosmological parameters, like \omegam\
and $\sigma_8$.  Our main results can be summarised as follows.

\begin{enumerate}
\item As predicted by analytical models \citep[see,
  e.g.,][]{matarrese00,loverde08}, the main effects in cluster counts
  due to the presence of some level of primordial non-Gaussianity are
  for high masses and redshifts. In particular, for haloes with
  $M_{200} > 10^{14} h^{-1} M_\odot$ at $z>1$, the differences with
  respect to the corresponding Gaussian models are about 20 per cent.

\item When the power spectrum normalization suggested by \wmapthree\ is
  adopted, models with a moderate level of non-Gaussianity (\fnl=$\pm
  100$) well reproduce the observed cluster counts derived from
  \rosat\ cluster survey.  However, the dependence on \fnl\ is very
  weak, when compared to the one on $\sigma_8$, which must be
  independently estimated to fully exploit cluster counts as a probe
  of primordial non-Gaussianity.

\item We predict the expected number and redshift distribution of the
  galaxy clusters that will be detected in two future cluster surveys:
  \erosita\ (X-ray) and \spt\ (SZ).  The effects due to a moderate
  primordial non-Gaussianity are in general of few per cent, reaching
  about 20 per cent only at high $z$.  In general, the fact that
  it is easier to detect high-$z$ objects with SZ observations, because 
  of the absence of cosmological dimming, makes \spt\ a more promising probe for
  obtaining constraints on \fnl.  However, once again,
  the results show a strong degeneracy between \fnl\ and other
  cosmological parameters.  Similar conclusions can be also drawn when
  analysing the power spectrum of the tSZ signal produced by galaxy
  clusters.

\item On the whole, the best strategy to detect the signatures of
  primordial non-Gaussianity in the LSS is to perform deep
  cluster surveys, together with suitable optical follow-ups for the
  determination of their redshifts. With this kind of 
  observational dataset, it
  would be possible to constrain $\sigma_8$ using low-redshift objects
  and analyse the $dN/dz$ in the range $0.5 \lesssim z \lesssim 1$ to
  constrain the value of \fnl. If with this method 
  future surveys will allow to reduce the uncertainties on $\sigma_8$ to 
  about 0.01, this would make possible to detect moderate 
  non-Gaussianities of the order of \fnl$=\pm100$ \citep[see also][]
  {sefusatti06,oguri09}.

\end{enumerate}

In conclusion, the results of this paper confirm the power of 
statistical tests based on galaxy clusters as a probe for primordial 
non-Gaussianity. In particular the detection of objects in the 
high-mass tail at sufficiently large redshift, as possible in future
SZ wide surveys like \spt, will be certainly useful to improve the 
constraints on \fnl\ coming from alternative methods, like CMB, ISW 
and galaxy biasing.

\section*{Acknowledgments}
Computations have been performed on the IBM-SP5 at CINECA (Consorzio
Interuniversitario del Nord-Est per il Calcolo Automatico), Bologna,
with CPU time assigned under an INAF-CINECA grant, and on the IBM-SP4
machine at the ``Rechenzentrum der Max-Planck-Gesellschaft'' at the
Max-Planck Institut fuer Plasmaphysik with CPU time assigned to the
``Max-Planck-Institut f\"ur Astrophysik'' and at the
``Leibniz-Rechenzentrum'' with CPU time assigned to the Project
``h0073''. We acknowledge the support of grant 
ANR-06-JCJC-0141 and the DFG cluster of
excellence Origin and Structure of the Universe.  We also 
acknowledge partial support by ASI contract I/016/07/0 ``COFIS'',
ASI-INAF I/023/05/0 and ASI-INAF I/088/06/0. We acknowledge useful 
discussions with S.Ameglio, S.Ettori and L.Verde.

\bibliographystyle{mn2e}
\newcommand{\noopsort}[1]{}

\label{lastpage}
\end{document}